\documentclass[floats,preprint,prb,aps]{revtex4}
\begin{document}

\vspace{0.5cm} \preprint{{\em Submitted to Phys. Rev. {\bf B}}
\hspace{3.0in} Web galley}
\date{\today}
\vspace{2.7in}

\title{Time-Dependent Quasiparticle Current Density Functional Theory of X-Ray Nonlinear  Response Functions}
\author{Oleg Berman$^{1}$ and Shaul Mukamel$^{2}$}
\affiliation{$^1$Department of Chemistry, University of Rochester,
Rochester, New York 14627-0216 \\
 $^2$Department of Chemistry,
University of California, Irvine, California 92697-2025}


\date{\today}

\vspace{2.7in}

\begin{abstract}

A real-space representation of the current response of
many-electron systems with possible applications to  x-ray
nonlinear spectroscopy and magnetic susceptibilities is developed.
Closed expressions for the linear, quadratic and third-order
response functions are derived by solving the adiabatic Time
Dependent Current Density Functional (TDCDFT) equations for the
single-electron density matrix in Liouville space.

\vspace{1.0 in}

PACS numbers: 71.15.Mb; 32.30.Rj; 71.45.Gm; 73.22.-f; 31.15.Lc

Key words: time dependent current density functional theory, x-ray
nonlinear spectroscopy, quasiparticles, exchange, correlation,
dielectric and magnetic response functions.

\end{abstract}

\maketitle {}

\section{Introduction}\label{intr}

Time Dependent Current Density Functional Theory (TDCDFT) offers a
computationally tractable framework for computing currents and
response functions of many-electron systems in response to
external electric and magnetic perturbations~\cite{VignaleKohn}.
The time-dependent linear paramagnetic susceptibilities are
calculated as the linear response of a noninteracting Kohn-Sham
reference system to an effective vector-potential, which consists
of the external field, together with the Hartree and the
exchange-correlation contributions~\cite{VignaleKohn}. For the
sake of computing current-related properties, it is natural to
consider the effective potential to be a functional of current and
charge density rather than charge density alone, as in standard
Time Dependent Density Functional Theory (TDDFT)~\cite{Gross}.
Another reason for applying TDCDFT to crystals is connected with
the recent argument that there is a one-to-one correspondence
between time-varying periodic potentials and the current density
but not with the charge density~\cite{Burke}. TDCDFT was
successfully used for calculating the polarizability of conjugated
polymers~\cite{Boeijprl,Boeij}. Current density functional theory
(CDFT) yields exact response functions~\cite{VignaleRasolt} to
static external potentials and TDCDFT is thus expected to provide
reasonable approximations for time-dependent current properties.

The linear magnetic susceptibility (the response of the current to
an external vector-potential) of the Kohn-Sham non-interacting
system has been calculated using the local current density
exchange correlation kernel for the electron
gas~\cite{VignaleKohn}. However, computing the response functions
of the interacting system (where the quasiparticle energies cannot
be expressed as differences between Kohn-Sham energy levels),
requires the solution of a chain of integral
equations~\cite{VignaleKohn}, whose computational cost rapidly
increases with the nonlinear order of the response. In this letter
we compute current response functions by recasting the TDCDDT
equations in terms of the reduced single electron density matrix
for an $N$ electron system $\rho (\mathbf{r}, \mathbf{r_{1}},t) =
\sum_{n=1}^{N}\psi _{n}(\mathbf{r}, t)\psi
_{n}^{*}(\mathbf{r_{1}}, t)$~\cite{Mukamel}, where
$\psi_{n}(\mathbf{r})$ are the Kohn-Sham orbitals. Closed
expressions are derived for the linear, quadratic and third-order
response functions~\cite{Gross,Berman,Chernyak} (the response of
the total polarization current to the external electric field) by
solving an eigenvalue equation in Liouville space. The
quasiparticle frequencies are not given simply as differences of
Kohn-Sham orbital energies~\cite{VignaleRasolt}. Current response
functions should also be particularly suitable for computing the
resonant nonlinear response to x-ray fields\cite{19}.  Due to the
short wavelength, the dipole approximation does not generally
hold, and x-ray susceptibilities may be expressed in terms of
multitime correlation functions of currents and charge densities
\cite{19,Tanaka}. TDCDFT thus provides a natural direct
computational approach for nonlinear x-ray spectroscopy.

\section{Time dependent Kohn-Sham current-density functional equations for the
single electron density matrix} \label{sec.ham1}

TDCDFT maps the original system of interacting electrons onto an
effective system of non-interacting electrons subjected to an
exchange-correlation scalar and vector
potentials~\cite{VignaleKohn}, constructed to yield the same
charge density and current profiles as the interacting system. The
Kohn-Sham TDCDFT equations of motion for the time-dependent
density matrix are
\begin{eqnarray} \label{eqroj}
i \frac{\partial \rho (\mathbf{r}, \mathbf{r_{1}},t)}{\partial t}
&=&   \left( \widehat H_{KS}^{0}(\mathbf{r},t) - \widehat
H_{KS}^{0*}(\mathbf{r}_{1},t) \right) \rho (\mathbf{r},
\mathbf{r_{1}},t) -
(\mathbf{j}(\mathbf{r},t)\mathbf{A}(\mathbf{r},t) -
\mathbf{j}^{*}(\mathbf{r}_{1},t)\mathbf{A}^{*}(\mathbf{r}_{1},t))\nonumber \\
&+&  \left(\frac{e^{2}}{2mc}A^{2}(\mathbf{r},t)n(\mathbf{r},t) -
\frac{e^{2}}{2mc}A^{2*}(\mathbf{r}_{1},t)n^{*}(\mathbf{r}_{1},t)
\right);
\end{eqnarray}
where $e(m)$ is the electron charge (mass), and we set $\hbar =
1$. The time-dependent charge density and paramagnetic electronic
current are given by
\begin{eqnarray} \label{roj}
 n(\mathbf{r},t) = \rho(\mathbf{r},\mathbf{r},t )  ;
\hspace{0.7in} \mathbf{j} (\mathbf{r},t) = - \frac{ie}{2m}\left[
(\nabla_{\mathbf r_{1}} - \nabla_{\mathbf r})\delta
\rho(\mathbf{r},\mathbf{r}_{1},t ) \right]_{\mathbf{r} =
\mathbf{r}_{1}}.
\end{eqnarray}
The observed (physical) current, which enters the continuity
equation, is given by
\begin{eqnarray} \label{tot_cur}
\mathbf{J} (\mathbf r, t) = \mathbf{j} (\mathbf r, t) -
\frac{e^{2}}{2mc}\mathbf{A}(\mathbf{r},t)n (\mathbf{r},t);
\end{eqnarray}
$\widehat H_{KS}^{0}$ is the Kohn-Sham Hamiltonian $ \widehat
H_{KS}^{0}(\mathbf{r},t) \equiv \widehat H_{KS}^{0}[n
(\mathbf{r},t), \mathbf{j} (\mathbf{r},t)]$. The remaining terms
in Eq.~(\ref{eqroj}) represent the coupling to an external
vector-potential.
\begin{eqnarray} \label{hks}
   \widehat H_{KS}^{0}[n (\mathbf{r},t), j (\mathbf{r},t)]
   = -\frac{1}{2m}\nabla_{\mathbf{r}}^{2} + U_{KS}(\mathbf{r},t),
\end{eqnarray}
where the potential $U_{KS}(\mathbf{r},t)= U_{KS}[n
(\mathbf{r},t), j (\mathbf{r},t)]$ is
\begin{eqnarray} \label{uks}
   U_{KS}(\mathbf{r},t)
   =  - \frac{e}{c}\nabla_{\mathbf{r}}\mathbf{A}_{xc}(\mathbf{r},t)
     - \frac{e}{c}\mathbf{A}_{xc}(\mathbf{r},t) \nabla_{\mathbf{r}}
     + U_{KS}^{0}(\mathbf{r},t) + U_{0}(\mathbf{r}).
\end{eqnarray}
The exchange-correlation vector potential $\mathbf{A}_{xc}[n
(\mathbf{r},t), j (\mathbf{r},t)]$  and the Kohn-Sham scalar
external potential $ U_{KS}^{0}[n (\mathbf{r},t), j
(\mathbf{r},t)]$ are functionals of both the charge density and
the current\cite{VignaleKohn}. The scalar potential is given by:
\begin{eqnarray} \label{uks0}
  U_{KS}^{0}(\mathbf{r},t) = \int d\mathbf{r_{1}}\frac{n(\mathbf{r}_{1},t)e^{2}}{|\mathbf{r}-\mathbf{r_{1}}|} +
U_{xc}(\mathbf{r},t) ;
\end{eqnarray}
$U_{0}(\mathbf{r})$ is the field created by nuclei and $U_{xc}[n
(\mathbf{r},t), j (\mathbf{r},t)](\mathbf{r})$ is the
exchange-correlation potential in the adiabatic approximation. The
time-dependent external potential is $U_{ext}(\mathbf{r}, t) =
U_{0}(\mathbf{r})$ at time $t\leq t_{0}$ and $U_{ext}(\mathbf{r},
t) = U_{0}(\mathbf{r})+ U_{1}(\mathbf{r}, t)$ for $t > t_{0}$.
$\mathbf{A}_{xc}$ adds a magnetic field induced by the
exchange-correlation interaction between electrons. Note that
unlike the paramagnetic canonical current, $\mathbf{J}$ is gauge
invariant~\cite{VignaleKohn}.

The stationary solution of Eq.~(\ref{eqroj}) gives the ground
state single electron density matrix $\bar{\rho}
(\mathbf{r},\mathbf{r_{1}})$ which carries no current. We then set
$\rho(\mathbf{r},\mathbf{r_{1}},t) \equiv \bar \rho
(\mathbf{r},\mathbf{r_{1}}) +
\delta\rho(\mathbf{r},\mathbf{r_{1}},t)$ where $\delta\rho$
represents the changes induced by $U_{1}(\mathbf{r}, t)$. Its
diagonal elements $\delta n \mathbf{(r)}= \delta \rho
(\mathbf{r},\mathbf{r},t)$ give the changes in charge density,
whereas the off-diagonal elements represent the changes in
electronic coherences between two points. The physical current may
be obtained by expanding $\delta\rho$ in powers of the external
vector potential $\mathbf{A}(\mathbf{r}, t)$: $\delta\rho = \delta
\rho _{1} + \delta\rho _{2} + ... $ and solving Eqs.~(\ref{eqroj})
and~(\ref{roj}) self-consistently for $\delta \rho$ order by
order. To that end we first recast $U_{KS}[n (\mathbf{r},t), j
(\mathbf{r},t)]$ as a functional of the paramagnetic current
$\mathbf{j}(\mathbf{r},t)$ alone. This is done by substituting the
total current $\mathbf{J}(\mathbf{r},t)$ (Eq.~(\ref{tot_cur}))
into the continuity relation between the charge density and the
total current
\begin{equation}
 \delta n(\mathbf{r},t) = -\frac{1}{e}\int_{0}^{t}\nabla_{\mathbf{r}}\mathbf{J}(\mathbf{r},\tau) d\tau .
  \label{cont}
\end{equation}
Solving Eq.~(\ref{cont}) self-consistently for the charge density
$n (\mathbf{r},t)$ in terms of the paramagnetic current
$\mathbf{j}(\mathbf{r},t)$, and substituting it in $U_{KS}[n
(\mathbf{r},t), j (\mathbf{r},t)]$ eliminates the explicit
dependence on the charge density. Expanding $U_{KS}$ around $\bar
\rho$ to second order in $\mathbf{j}$, we obtain
\begin{eqnarray} \label{taylor}
&& U_{KS}[\mathbf{j}(\mathbf{r},t)](\mathbf{r}) = U_{KS}^{0}[\bar
\rho](\mathbf{r}) + \int  d\mathbf{r}_{1} \left(
\frac{e^2}{|\mathbf r - \mathbf r_{1}|} + \tilde{f_{xc}}[\bar
\rho] (\mathbf r, \mathbf r_{1}) \right)
\mathbf{j} (\mathbf{r}_{1},t)  \nonumber \\
 &+&  \int d\mathbf{r}_{1}
\int d\mathbf{r}_{2} \tilde{g_{xc}}[\bar \rho] (\mathbf r, \mathbf
r_{1}, \mathbf{r}_{2}) \mathbf{j} (\mathbf{r}_{1},t) \mathbf{j}
(\mathbf{r}_{2},t);
\end{eqnarray}
where $\tilde{f_{xc}}[\bar \rho] (\mathbf r, \mathbf r_{1})$ and
$\tilde{g_{xc}}[\bar \rho] (\mathbf r, \mathbf r_{1}, \mathbf
r_{2})$ are the first and the second  order adiabatic
exchange-correlation kernels. We have made the commonly used
adiabatic approximation where we assume that the kernels are
time-independent.

\section{Quasiparticle representation of X-Ray Nonlinear  Response Functions} \label{sec.ham}

We next separate $\delta\rho$ into an electron-hole, interband,
($\xi$) and intraband ($T$) components $\delta\rho
(\mathbf{r},\mathbf{r_{1}},t) = \xi(\mathbf{r},\mathbf{r_{1}},t) +
T (\xi(\mathbf{r},\mathbf{r_{1}},t))$. It follows from the
idempotent property of $\rho$, that $T$ is uniquely determined by
$\xi$~\cite{Chernyak}. The matrix elements of $\xi$, unlike those
of $\delta\rho$, constitute independent variables, that can be
used to construct a quasiparticle representation.

The quasiparticle spectrum is obtained by solving the linearized
Kohn-Sham eigenvalue equation
\begin{eqnarray}
\label{eigenvalue}
 L \xi_\alpha (\mathbf r, \mathbf r_{1}) =
\Omega_\alpha \xi_\alpha (\mathbf r, \mathbf r_{1}) ,
\end{eqnarray}
where
 \begin{eqnarray}
 \label{Lf}
L \xi_\alpha (\mathbf r, \mathbf r_{1}) &=&
-\frac{(\nabla_{\mathbf{r}}^{2} -
\nabla_{\mathbf{r_{1}}}^{2})\xi_\alpha (\mathbf r, \mathbf
r_{1})}{2m} +  \bar \rho (\mathbf{r},\mathbf{r}_{1})\left(\int
d\mathbf{r}_{1} \int d\mathbf{r}_{2} f_{xc}'[\bar \rho] (\mathbf
r, \mathbf r_{1}, \mathbf r_{2})\xi_\alpha (\mathbf r_{1}, \mathbf
r_{2}) \right) \nonumber \\
&-&  \bar \rho ^{*}(\mathbf{r}_{1},\mathbf{r}) \left(\int
d\mathbf{r} \int d\mathbf{r}_{2} f_{xc}'^{*}[\bar \rho] (\mathbf
r_{1}, \mathbf r, \mathbf r_{2})\xi_\alpha (\mathbf r_{2}, \mathbf
r) \right)  ,
  \end{eqnarray}
 \begin{eqnarray}
 \label{f''}
f_{xc}'[\bar \rho] (\mathbf r, \mathbf r_{1}, \mathbf r_{2}) =
\tilde{f_{xc}}'[\bar \rho] (\mathbf r, \mathbf
r_{1})\delta(\mathbf r_{1}- \mathbf r_{2})\hat{v}(\mathbf r_{1},
\mathbf r_{2}),
  \end{eqnarray}
and $\hat{v}$ is the velocity operator
 \begin{eqnarray}
 \label{vel}
 \hat{v}(\mathbf r,\mathbf
r_{1}) = - \frac{i}{2m}(\nabla_{\mathbf r_{1}} - \nabla_{\mathbf
r}).
 \end{eqnarray}

The quasiparticle modes $\xi_{\alpha}$ come in pairs $\alpha = \pm
1, \pm 2, \ldots$ with $\Omega_{-\alpha} \equiv \Omega_\alpha$.
Their orthonormality and algebraic properties make it possible to
expand an arbitrary interband matrix~\cite{Chernyak} in the form
 \begin{eqnarray}
 \label{eqosc6}
\xi (\mathbf r, \mathbf r_{1}, t)= \sum_{\alpha}
\xi_{\alpha}(\mathbf r, \mathbf r_{1}) z_{\alpha}(t),
 \end{eqnarray}
where $z_{\alpha}(t) = \langle \xi_\alpha^\dagger |\xi
(t)\rangle$, where the scalar product of any two interband
matrices $\xi$ and $\eta$ is defined by\cite{Berman}
\begin{equation}
\label{scalar}
 \langle \xi|\eta \rangle \equiv \int d\mathbf{r} \int d\mathbf{r'}\bar
\rho[\xi^\dagger,\eta](\mathbf{r},\mathbf{r}')\delta(\mathbf{r} -
\mathbf{r}').
 \end{equation}
The bra (ket) notation underscores the similarity with Dirac's
Hilbert space notation. $z_{-\alpha}(t) =z_{\alpha}^*(t)$
constitute complex oscillator amplitudes. We shall denote their
perturbative expansion in the external vector-potential
$\mathbf{A}(\mathbf{r},t)$ by $z_\alpha^{(1)},
z_\alpha^{(2)}\cdots$.

The paramagnetic current is calculated by taking the expectation
value of the velocity $\hat{v}(\mathbf r,\mathbf r_{1})$ with
respect to the time-dependent density matrix
\begin{eqnarray} \label{eqcur3}
\mathbf{j}(\mathbf r, t) &=& \sum_{\alpha} \mathbf{
j}_{\alpha}(\mathbf r) z_{\alpha}(t) + \frac {1}{2} \sum_{\alpha
\beta } \mathbf{j}_{\alpha \beta}(\mathbf r) z_{\alpha}(t)
z_{\beta}(t) + \nonumber \\
&+& \frac {1}{3} \sum_{\alpha \beta \gamma} \mathbf{j}_{\alpha,
\beta \gamma}(\mathbf r,\mathbf r')z_{\alpha}(t)
z_{\beta}(t)z_{\gamma}(t),
\end{eqnarray}
where ($\alpha, \beta, \gamma = \pm 1, \pm 2, \ldots $) and we
only retained terms that contribute to the second order response.
Here

\begin{eqnarray} \label{3.00a}
\mathbf{j}_{\alpha}(\mathbf r) = - \frac{ie}{2m} \int d \mathbf
r_{1} \left[ \delta (\mathbf r - \mathbf r_{1})(\nabla_{\mathbf
r_{1}} - \nabla_{\mathbf r}) \xi_{\alpha}(\mathbf r, \mathbf
r_{1}) \right],
\end{eqnarray}
\begin{eqnarray} \label{3.00aa}
 \mathbf{j}_{\alpha \beta} (\mathbf r) &=& -
\frac{ie}{2m} \int d \mathbf r_{1} \left[\delta (\mathbf r -
\mathbf r_{1})(\nabla_{\mathbf r_{1}} - \nabla_{\mathbf r}) \int
d\mathbf{r}_{2}\left((\delta(\mathbf{r}_{2} - \mathbf{r}) \right.
\right. \nonumber \\ &-& \left. \left. 2 \bar n(\mathbf r, \mathbf
r_{2}))\int d\mathbf{r}_{3}(\xi_{\alpha}(\mathbf r_{2}, \mathbf
r_{3}) \xi_{\beta}(\mathbf r_{3}, \mathbf r_{1}) +
\xi_{\beta}(\mathbf r_{2}, \mathbf r_{3})
 \xi_{\alpha}(\mathbf r_{3}, \mathbf r_{1}))\right) \right],
\end{eqnarray}
\begin{eqnarray} \label{3.00aaa}
 \mathbf{j}_{\alpha \beta \gamma} (\mathbf r) &=&
\frac{ie}{2m} \int d \mathbf r_{1} \left[\delta (\mathbf r -
\mathbf r_{1})(\nabla_{\mathbf r_{1}} - \nabla_{\mathbf r}) \int
d\mathbf{r}_{2}\left(\xi_{\alpha}(\mathbf r, \mathbf r_{2})
\right. \right. \nonumber \\ &&  \left. \left . \int
d\mathbf{r}_{3}(\xi_{\alpha}(\mathbf r_{2}, \mathbf r_{3})
\xi_{\beta}(\mathbf r_{3}, \mathbf r_{1}) + \xi_{\beta}(\mathbf
r_{2}, \mathbf r_{3})
 \xi_{\alpha}(\mathbf r_{3}, \mathbf r_{1}))\right) \right].
\end{eqnarray}
The collective electronic oscillator (CEO) expansion for the
charge density $n (\mathbf r, t)$ is given by~\cite{Berman}
\begin{eqnarray} \label{eqpolar22}
\delta n (\mathbf r, t) &=& \sum_{\alpha} n_{\alpha}(\mathbf r)
z_{\alpha}(t) + \frac {1}{2} \sum_{\alpha \beta } n_{\alpha
\beta}(\mathbf r)
z_{\alpha}(t) z_{\beta}(t) \nonumber \\
&+& \frac {1}{3} \sum_{\alpha \beta \gamma} n_{\alpha, \beta
\gamma}(\mathbf r)z_{\alpha}(t) z_{\beta}(t)z_{\gamma}(t),
\nonumber \\ && \alpha,  \beta, \gamma = \pm 1, \pm 2, \ldots
\end{eqnarray}
The coefficients of this expansion are given by Eqs.~(29)-~(31) in
Ref.\cite{Berman}.

 The equations of motion for $\xi$ can be
obtained from Eq.~(\ref{eqroj}) by expressing the density matrix
in terms of $\xi$ and $T(\xi)$~\cite{Berman}. Equations of motions
for $z^{(1)}(t)$ are derived in terms of $\mathbf{A}(\mathbf r,t)$
and $\mathbf{j}(\mathbf r)$ by substituting the mode expansion of
$\xi$ into these equations. Substituting the solution $z^{(1)}(t)$
into Eq.(\ref{eqcur3}), gives
\begin{eqnarray} \label{restcur11}
j^{(1)}_{\lambda_{s}} (\mathbf r, t) = \int_{-\infty}^t d \tau
\int d\mathbf{r_{1}} \sum_{\mu = x, y, z}
\tilde{\chi}_{\lambda_{1} \lambda_{s}}^{(1)} (t, \tau,\mathbf r,
\mathbf r_{1}) A_{\lambda_{1}} (\mathbf r_{1},\tau) ,
\end{eqnarray}
where $\tilde{\chi}^{(1)}$ is the linear paramagnetic
susceptibility, and $\lambda_{s}, \lambda_{1}$ are Cartesian
tensor components.

We further introduce the observed  susceptibility $\chi_{\lambda
\mu}^{(1)} (t, \tau,\mathbf r, \mathbf r_{1}) $ defined by
replacing the paramagnetic current $\mathbf{j} (\mathbf r, t) $
with the physical current $\mathbf{J} (\mathbf r, t)$ in
Eq.~(\ref{restcur11}). Substituting $\mathbf{j}^{(1)} (\mathbf r,
t)$ from Eq.~(\ref{tot_cur}) into Eq.~(\ref{restcur11}), we obtain
\begin{eqnarray} \label{sus}
 \chi_{\lambda
\mu}^{(1)} (t, \tau,\mathbf r, \mathbf r_{1}) =
\tilde{\chi}_{\lambda \mu}^{(1)} (t, \tau,\mathbf r, \mathbf
r_{1}) - \frac{1}{mc}\bar{n}(\mathbf{r}_{1})\delta(\mathbf r_{1} -
\mathbf r)\delta(\tau - t)\delta_{\lambda\mu} .
\end{eqnarray}
We next derive equations of motions for $z^{(1)}(t)$ in terms of
$\mathbf{A}(\mathbf r,t)$ and $\mathbf{j}_{\alpha}(\mathbf r)$, by
substituting the CEO expansion of $\xi$ from Eq.~(\ref{eqosc6})
into the equation of motion for $\xi$, which can be obtained from
Eq.~(\ref{eqroj}) by expansion of the density matrix on $\xi$ and
$T(\xi)$~\cite{Berman}. Solving these equations we substitute
$z^{(1)}(t)$ into Eq.(\ref{eqcur3}).
\begin{eqnarray} \label{con1}
\tilde{\chi}_{\lambda \mu}^{(1)}(\omega,\mathbf r, \mathbf r_{1})
& = &   \sum_{\alpha = 1, 2, \ldots} \frac{2 \Omega_\alpha
j_{\alpha\lambda} (\mathbf r)j_{\alpha\mu} (\mathbf r_{1})}
{\Omega_\alpha^2 - \omega^2}.
\end{eqnarray}

Optical with x-ray signals are most conveniently expressed using
response functions which connect the polarization with the
electric field.  For example, the linear response function $
\sigma_{\lambda_{1} \lambda_{s}}^{(1)} (t, \tau,\mathbf r, \mathbf
r_{1})$ to first order in the external field $E(\mathbf{r},t)$ is
defined as:
\begin{eqnarray} \label{restcur1}
P^{(1)}_{\lambda_{s}} (\mathbf r, t) = \int_{-\infty}^t d \tau
\int d\mathbf{r_{1}} \sum_{\mu = x, y, z}  \sigma_{\lambda_{1}
\lambda_{s}}^{(1)} (t, \tau,\mathbf r, \mathbf r_{1})
E_{\lambda_{1}} (\mathbf r_{1},\tau),
\end{eqnarray}
where $\mathbf{P}(\mathbf r, t)$ is the total
polarization\cite{Tanaka}. $\sigma^{(1)} (t, \tau,\mathbf r,
\mathbf r_{1})$ can be obtained from Eq.~(\ref{restcur11}) by
noting that $\mathbf{j}$ is connected to $\mathbf{J}$ through
Eq.~(\ref{tot_cur}); $\mathbf{J}$ is connected to $\mathbf{P}$
through\cite{Tanaka} $\mathbf{P}(\mathbf r, t) =
\int_{-\infty}^{t}d\tau \mathbf{J}(\mathbf r, \tau)$ and
$\mathbf{A}(\mathbf{r},\omega)= -i c \mathbf{E} (\mathbf
{r},\omega)/\omega$. Using these relations we obtain\cite{Tanaka}
\begin{eqnarray} \label{rel}
\sigma^{(n)} (\omega,\mathbf r,\mathbf r_{n},\ldots ,\mathbf
r_{1}, \omega_{n},\ldots , \omega_{1})  =
\frac{i^{1-n}}{\omega_{1}\omega_{2}\ldots
\omega_{n}\omega}\chi^{(n)} (\omega,\mathbf r,\mathbf r_{n},\ldots
,\mathbf r_{1}, \omega_{n},\ldots , \omega_{1}),
\end{eqnarray}
this gives for the linear response
\begin{eqnarray} \label{con1b}
&& \sigma_{\lambda_{1} \lambda_{s}}^{(1)} (\omega,\mathbf r,
\mathbf r_{1}) = \frac{1}{\omega^{2}} \chi_{\lambda_{1}
\lambda_{s}}^{(1)} (\omega = \omega_{1},\mathbf
r, \mathbf r_{1}) \nonumber \\
 && = \frac{1}{\omega^{2}}
2\pi \left( \sum_{\alpha = \pm 1, \pm 2, \ldots} \frac{2
\Omega_\alpha j_{\alpha}^{\lambda_{s}} (\mathbf
r)j_{\alpha}^{\lambda_{1}} (\mathbf r_{1})} {\Omega_\alpha^2 -
\omega^2} - \frac{e^{2}}{2mc}\bar{n}(\mathbf{r}_{1})\delta(\mathbf
r_{1} - \mathbf r)\delta_{\lambda_{s}\lambda_{1}} \right),
\end{eqnarray}
where the first term in the bracket is $\tilde{\chi}_{\lambda_{s}
\lambda_{1}}^{(1)}(\omega, \mathbf r, \mathbf r_1)$, and $\bar{n}
(\mathbf r)$ is the ground state charge density. Eq.~(\ref{con1b})
provides a microscopic algorithm for computing the Kubo
formula~\cite{Haug}; all quantities are obtained from the
quasiparticle modes.

To calculate the second order response function $\sigma^{(2)}$
\begin{eqnarray} \label{restcur2}
P_{\lambda_{s}}^{(2)} (\mathbf r, t) &=&
\frac{1}{2}\int_{-\infty}^t d \tau_1 \int_{-\infty}^{t} d \tau_2
\int d\mathbf{r_{1}}\int d\mathbf{r_{2}} E _{\lambda_{1}}(\mathbf
r_{1},\tau_1) E_{\lambda_{2}} (\mathbf r_{2},\tau_2) \nonumber \\
&& \sigma_{\lambda_{1} \lambda_{2}\lambda_{s}}^{(2)} (t, \tau_1,
\tau_2,\mathbf r, \mathbf r_{1}, \mathbf r_{2}),
\end{eqnarray}
we introduce the second-order exchange-correlation kernel
$g_{xc}$, obtained by expanding the exchange correlation potential
$U_{KS}[j(\mathbf{r},t)](\mathbf{r})$ to the second order by
$\delta \mathbf{j}$ (Eq.~(\ref{taylor}))
\begin{eqnarray}
\label{g''} g_{xc}[\bar n] (\mathbf r, \mathbf r_{1}, \mathbf
r_{2},\mathbf r_{2}',\mathbf{r}_{4}) = \tilde{g_{xc}}[\bar n]
(\mathbf r, \mathbf r_{1}, \mathbf{r}_{3})\delta(\mathbf r_{1}-
\mathbf r_{2})\delta(\mathbf r_{2}'- \mathbf r_{4})\hat{v}(\mathbf
r_{1}, \mathbf r_{2})\hat{v}(\mathbf r_{2}', \mathbf r_{4}).
\end{eqnarray}
Repeating the procedure used for $\sigma^{(1)}$ to the next order
we obtain the second order paramagnetic susceptibility
\begin{eqnarray}\label{paramagnetic}
&& \tilde{\chi}_{\lambda_{1} \lambda_{2}\lambda_{s}
}^{(2)}(\omega_1, \omega_2,\mathbf r, \mathbf r_{1}, \mathbf
r_{2}) = - 2 \sum_{\alpha \lambda_{s}\beta\gamma}
\frac{V_{g(-\alpha
\beta\gamma)}(\mathbf{r},\mathbf{r}_{1},\mathbf{r}_{2})j_{\alpha}^{\lambda_{s}}(\mathbf
r) j_{-\beta}^{\lambda_{1}}(\mathbf r_{1})
j_{-\gamma}^{\lambda_{2} }(\mathbf r_{2}) s_{\alpha }
s_{\beta}}{(\Omega_{\alpha } - \omega_1
- \omega_2)(\Omega_{\beta} - \omega_1)(\Omega_{\gamma} - \omega_2)} \nonumber \\
&+& \sum_{\alpha\beta} \frac{j_{-\alpha \beta}^{\lambda_{s}
}(\mathbf r)j_{\alpha}^{\lambda_{1}}(\mathbf r_{1})
j_{-\beta}^{\lambda_{2}}(\mathbf r_{2}) s_{\alpha}
s_{\beta}}{(\Omega_{\alpha} - \omega_1 - \omega_2)(\Omega_{\beta}
- \omega_1)} + \sum_{\alpha\beta} \frac{j_{-\alpha \beta}^{
\lambda_{s} }(\mathbf r)j_{\alpha}^{\lambda_{1}}(\mathbf r_{1})
j_{-\beta}^{\lambda_{2}}(\mathbf r_{2}) s_{\alpha}
s_{\beta}}{(\Omega_{\alpha} - \omega_1
- \omega_2)(\Omega_{\beta} - \omega_2)} \nonumber \\
&+& \sum_{\alpha \beta} \frac{j_{\alpha \beta}^{\lambda_{s}
}(\mathbf r)j_{\alpha}^{\lambda_{1}}(\mathbf r_{1})
j_{-\beta}^{\lambda_{2}}(\mathbf r_{2}) s_{\alpha}
s_{\beta}}{(s_{\alpha} \Omega_{\alpha} - \omega_1)(s_{\beta}
\Omega_{\beta} - \omega_2)}, \hspace{1em} \alpha, \beta, \gamma =
\pm 1, \pm2, \ldots, \label{con2}
\end{eqnarray}
where $s_{\alpha} \equiv sign(\alpha)$.
$V_{g(-\alpha\beta\gamma)}$ is obtained by substituting the
exchange-correlation kernels $f_{xc}$ from Eq.~(\ref{f''}) and
$g_{xc}$ into the expression for $V_{g(-\alpha\beta\gamma)}$ given
in Ref.\cite{Berman}.

Similar to the  linear response function, the second-order
response function is finally obtained by expanding of the charge
density $n(\mathbf{r},t)$ in the  modes (Eq.~(\ref{eqpolar22})):
\begin{eqnarray} \label{con2b}
&& \sigma_{\lambda_{1} \lambda_{2} \lambda_{s}
}^{(2)}(\omega_{1},\omega_{2},\mathbf r, \mathbf r_{1}, \mathbf
r_{2}) = - \frac{i}{\omega_{1} \omega_{2} (\omega_{1} +
\omega_{2})} 2\pi \left[ \tilde{\chi}_{\lambda_{s} \lambda_{1}
\lambda_{2}}^{(2)}(\omega_1, \omega_2,\mathbf r, \mathbf r_{1},
\mathbf
r_{2})   \right. \nonumber \\
&-& \left. \frac{e^{2}}{2mc} \left( \sum_{\alpha } \frac{2
\Omega_\alpha j_{\alpha}^{\lambda_{s}} (\mathbf r)n_{\alpha}
(\mathbf r_{1})} {\Omega_\alpha^2 - (\omega_{1} +
\omega_{2})^2}\delta(\mathbf r_{2}
- \mathbf r_{1})\delta_{\lambda_{2} \lambda_{1}} \right. \right. \nonumber \\
&& \left. \left. + \sum_{\alpha } \frac{2 \Omega_\alpha n_{\alpha}
(\mathbf r)j_{\alpha}^{\lambda_{1}} (\mathbf r_{1})}
{\Omega_\alpha^2 - (\omega_{1} + \omega_{2})^2}\delta(\mathbf r -
\mathbf r_{2})\delta_{ \lambda_{2} \lambda_{s}}\right)\right] .
\end{eqnarray}
 Higher response functions can be computed
similarly~\cite{Berman}. The third-order response function is
given in Appendix \ref{ap.2pz3}.

\section{Discussion}\label{disc}

To get the high-order paramagnetic susceptibilities in the
standard Hilbert space TDDCDT approach one needs to solve
self-consistently a chain of integral equations for each
order~\cite{VignaleKohn}. The linear paramagnetic susceptibility
in the standard Hilbert space TDDCDT approach is given by
Eqs.~(8)-~(9) in Ref.~\cite{VignaleKohn}. In contrast, the closed
expressions for the linear (Eq.~(\ref{con1})), second-order
(Eq.~(\ref{paramagnetic})) and the third-order
(Eq.~(\ref{paramagnetic11})) susceptibilities derived in this
paper use the CEO representation in Liouville space.

Correlation-function expressions for the linear, second-order and
third-order x-ray response functions were derived in
Eqs.(B1),(B2),(B3a)-~(B3d) in Ref.~\cite{Tanaka}.
Eqs.~(\ref{con1b}),~(\ref{con2b}) and~(\ref{con3b}) express these
TDCDFT response functions in the CEO representation, and provide a
computational scheme for nonlinear x-ray response functions.

TDDFT exchange-correlation functionals are better developed and
more widely used than their TDCDFT counterparts. TDDFT currents
can be obtained by simply modifying Eqs.~(\ref{f''}) by setting
$\hat{v} = 1$, and using the TDDFT exchange-correlation
kernels\cite{Gross,Berman} where the scalar exchange-correlation
potential depends only on charge density.

Finally we note that this work can be extended to include
non-adiabatic exchange-correlation potentials, as outlined
recently for the linear response~\cite{Chernyak2}. In general, the
exchange-correlation potential and exchange-correlation kernels
are time-dependent~\cite{GrossKohn}. This time dependence has been
neglected within the adiabatic approximation used here. If we
relax this approximation, the eigenvalue equation for the
Liouville superoperator $L$, Eq.~(\ref{eigenvalue}), should be
replaced by~\cite{Chernyak2}
\begin{eqnarray}
\label{eigenvalue2}
 L(\Omega_{\alpha}) \xi_\alpha (\mathbf r, \mathbf r') =
\Omega_\alpha \xi_\alpha (\mathbf r, \mathbf r') .
\end{eqnarray}
Methods for solving Eq.(\ref{eigenvalue2}) using the
frequency-dependent functional of Gross and Kohn~\cite{GrossKohn}
were described in~\cite{Chernyak2}.

\section*{Acknowledgements}

This article is based upon work supported by the National Science
Foundation under grant number (CHE-0132571). This support is
gratefully acknowledged.

\appendix

\section{The third-order response} \label{ap.2pz3}

For the third-order response function we obtain
\begin{eqnarray} \label{con3b}
&& \sigma_{\lambda_{1} \lambda_{2} \lambda_{3} \lambda_{s}
}^{(3)}(\omega_{1},\omega_{2},\omega_{3},\mathbf r, \mathbf r_{1},
\mathbf r_{2},\mathbf r_{3}) = - \frac{
1}{\omega_{1}\omega_{2}\omega_{3}(\omega_{1} + \omega_{2} +
\omega_{3})}2\pi \nonumber \\
&&\left[ \tilde{\chi}_{\lambda_{1} \lambda_{2} \lambda_{3}
\lambda_{s} }^{(3)}(\omega_{1},\omega_{2},\omega_{3},\mathbf r,
\mathbf r_{1}, \mathbf r_{2},\mathbf r_{3})   \right. \nonumber \\
&+& \left. \frac{e^{2}}{2mc} \left( F[j^{\lambda_{s}} (\mathbf r)n
(\mathbf r_{3})j^{\lambda_{1}}(\mathbf{r}_{1})]\delta(\mathbf
r_{2} - \mathbf r_{3})\delta_{\lambda_{2} \lambda_{3}}  +
F[j^{\lambda_{s}} (\mathbf r)n (\mathbf
r_{3})j^{\lambda_{1}}(\mathbf{r}_{1})]\delta(\mathbf r_{2} -
\mathbf r_{3})\delta_{\lambda_{2} \lambda_{3}} \right. \right.
\nonumber
\\ &+& \left. \left. F[n (\mathbf r)j^{\lambda_{2}}(\mathbf
r_{3})j^{\lambda_{1}}(\mathbf r_{1})]\delta(\mathbf r - \mathbf
r_{3})\delta_{\lambda_{2}
\lambda_{1}}\delta_{\lambda_{3} \lambda_{s}} \right) \right. \nonumber \\
&+& \left. \left(\frac{e^{2}}{2mc}\right)^{2}R^{(1)}(\omega_{1} +
\omega_{2} + \omega_{3},\mathbf r, \mathbf r_{2})\delta(\mathbf r
- \mathbf r_{3})\delta(\mathbf r_{2} - \mathbf
r_{1})\delta_{\lambda_{3} \lambda_{s}} \delta_{\lambda_{2}
\lambda_{1}}\right] ,
\end{eqnarray}
where $R^{(1)}(\omega_{1} + \omega_{2} + \omega_{3},\mathbf r,
\mathbf r_{2})$ is the linear density-density response given by
Eq.~(46) in Ref.\cite{Berman}; $\tilde{\chi}_{\lambda_{1}
\lambda_{2} \lambda_{3} \lambda_{s}
}^{(3)}(\omega_{1},\omega_{2},\omega_{3},\mathbf r, \mathbf r_{1},
\mathbf r_{2},\mathbf r_{3})$ is the third-order paramagnetic
susceptibility given by the r.h.s. of Eqs.~(C23)-~(C31) in
Ref.\cite{Berman} by replacing $\rho$ by
$j$:\\
\begin{equation} \label{pol3}
\tilde{\chi}_{\lambda_{1} \lambda_{2} \lambda_{3} \lambda_{s}
}^{(3)} (\omega_1, \omega_2, \omega_3, \mathbf r, \mathbf r',
\mathbf r'',\mathbf r''') = \sum_{\omega_1 \omega_2
\omega_3}^{perm} \left( \tilde{\chi}_{I}^{(3)} +
\tilde{\chi}_{II}^{(3)} + \tilde{\chi}_{III}^{(3)} + \ldots
\tilde{\chi}_{VIII}^{(3)} \right ),
\end{equation}
where
\begin{eqnarray} \label{pol3I}
&& \tilde{\chi}_{I}^{(3)} = \sum_{\alpha \beta \gamma}
\frac{j_{-\alpha\beta}^{\lambda_{s}}(\mathbf r) j_{-\beta
\gamma}^{\lambda_{1}}(\mathbf r') j_{\alpha}^{\lambda_{2}}(\mathbf
r'') j_{-\gamma}^{\lambda_{3}}(\mathbf r''') s_{\alpha} s_{\beta}
s_{\gamma}}{(\Omega_{\alpha} - \omega_1 - \omega_2
- \omega_3) (\Omega_{\beta} - \omega_2 - \omega_3) (\Omega_{\gamma} - \omega_3)},\\
\label{pol3II}
&& \tilde{\chi}_{II}^{(3)} = \sum_{\alpha \beta \gamma \delta}
\frac{j_{-\alpha \beta}^{\lambda_{s}}(\mathbf r) V_{g(-\beta
\gamma \delta)} j_{\alpha}^{\lambda_{1}}(\mathbf r')
j_{-\gamma}^{\lambda_{2}}(\mathbf r'')
j_{-\delta}^{\lambda_{3}}(\mathbf r''') s_{\alpha} s_{\beta}
s_{\gamma} s_{\delta}}{(\Omega_{\alpha} - \omega_1 - \omega_2 -
\omega_3)(\Omega_{\beta} - \omega_2
- \omega_3)(\Omega_{\gamma} - \omega_2)(\Omega_{\delta} - \omega_3)},\\
\label{pol3III}
&& \tilde{\chi}_{III}^{(3)} = \sum_{\alpha \beta \gamma}
\frac{j_{-\alpha \beta \gamma}^{\lambda_{s}}(\mathbf r)
j_{\alpha}^{\lambda_{1}}(\mathbf r')
j_{-\beta}^{\lambda_{2}}(\mathbf r'')
j_{-\gamma}^{\lambda_{3}}(\mathbf r''') s_{\alpha} s_{\beta}
s_{\gamma}}{(\Omega_{\alpha} -\omega_1 - \omega_2
- \omega_3)(\Omega_{\beta} - \omega_2 - \omega_3)(\Omega_{\gamma} - \omega_3)},\\
\label{pol3IV}
&& \tilde{\chi}_{IV}^{(3)} =  \sum_{\alpha \beta \gamma \delta}
\frac{2V_{g(-\alpha \beta \gamma)} j_{-\gamma
\delta}^{\lambda_{s}}(\mathbf r) j_{\alpha}^{\lambda_{1}}(\mathbf
r') j_{-\beta}^{\lambda_{2}}(\mathbf r'')
j_{-\delta}^{\lambda_{3}}(\mathbf r''') s_{\alpha} s_{\beta}
s_{\gamma} s_{\delta}}{(\Omega_{\alpha} -\omega_1 - \omega_2 -
\omega_3)(\Omega_{\beta} - \omega_1)(\Omega_{\gamma} - \omega_2 -
\omega_3)(\Omega_{\delta} - \omega_3)},\\ \label{pol3V}
&& \tilde{\chi}_{V}^{(3)} = \nonumber \\
&& \sum_{\alpha \beta \gamma \delta \eta} \frac{2V_{g(-\alpha\beta
\gamma)} V_{g(-\gamma \delta \eta)}
j_{\alpha}^{\lambda_{s}}(\mathbf r)
j_{-\beta}^{\lambda_{1}}(\mathbf r')
j_{-\delta}^{\lambda_{2}}(\mathbf r'')
j_{-\eta}^{\lambda_{3}}(\mathbf r''') s_{\alpha} s_{\beta}
s_{\gamma} s_{\delta} s_{\eta}}{(\Omega_{\alpha} -\omega_1 -
\omega_2 - \omega_3)(\Omega_{\beta} - \omega_1)(\Omega_{\gamma}
-\omega_2 -
\omega_3)(\Omega_{\delta} - \omega_2)(\Omega_{\eta} - \omega_3)}\\
\label{pol3VI}
&& \tilde{\chi}_{VI}^{(3)} = \sum_{\alpha \beta \gamma \delta}
\frac{V_{h(-\alpha \beta \gamma \delta)}
j_{\alpha}^{\lambda_{s}}(\mathbf r)
j_{-\beta}^{\lambda_{1}}(\mathbf r')
j_{-\gamma}^{\lambda_{2}}(\mathbf r'')
j_{-\delta}^{\lambda_{3}}(\mathbf r''') s_{\alpha} s_{\beta}
s_{\gamma} s_{\delta}}{(\Omega_{\alpha} -\omega_1 - \omega_2 -
\omega_3)(\Omega_{\beta}
- \omega_1)(\Omega_{\gamma} - \omega_2)(\Omega_{\delta} - \omega_3)},\\
\label{pol3VII}
&& \tilde{\chi}_{VII}^{(3)} = \sum_{\alpha \beta \gamma}
\frac{j_{\alpha\beta}^{\lambda_{s}}(\mathbf r) j_{-\beta
\gamma}^{\lambda_{1}}(\mathbf r')
j_{-\alpha}^{\lambda_{2}}(\mathbf r'')
j_{-\gamma}^{\lambda_{3}}(\mathbf r''') s_{\alpha} s_{\beta}
s_{\gamma}}{(\Omega_{\alpha} - \omega_1)( \Omega_{\beta}
- \omega_2 - \omega_3)(\Omega_{\gamma} - \omega_3)},\\
\label{pol3VIII}
&& \tilde{\chi}_{VIII}^{(3)} = \sum_{\alpha \beta \gamma \delta}
\frac{j_{\alpha \beta}^{\lambda_{s}}(\mathbf r) V_{g(-\beta \gamma
\delta)} j_{-\alpha}^{\lambda_{1}}(\mathbf r')
j_{-\gamma}^{\lambda_{2}}(\mathbf r'')
j_{-\delta}^{\lambda_{3}}(\mathbf r''') s_{\alpha} s_{\beta}
s_{\gamma} s_{\delta}}{(\Omega_{\alpha} - \omega_1)(\Omega_{\beta}
- \omega_2 - \omega_3)(\Omega_{\gamma} - \omega_2)(\Omega_{\delta}
- \omega_3)}.
\end{eqnarray}
Here $\nu = \alpha, \beta, \gamma, \delta, \eta = \pm 1, \pm 2,
\ldots $ and $\Omega_\nu$ is positive (negative) for all $\nu>0$
($\nu<0$) according to the convention $\Omega_{-\nu}=-\Omega_\nu$.

$F[j^{\lambda_{s}} (\mathbf r)n (\mathbf
r_{3})j^{\lambda_{1}}(\mathbf{r}_{1})]$ is determined by the
r.h.s. of Eq.~(\ref{con2b}) by replacing
$j^{\lambda_{1}}(\mathbf{r}_{1})$ by $n (\mathbf r_{3})$:
\begin{eqnarray} \label{con2bb}
&& F[j^{\lambda_{s}} (\mathbf r) n (\mathbf
r_{3})j^{\lambda_{1}}(\mathbf{r}_{1})] = - \frac{i}{\omega_{1}
\omega_{2} (\omega_{1} + \omega_{2})} 2\pi\left[
\hat{\chi}_{\lambda_{s} \lambda_{1} \lambda_{2}}^{(2)}(\omega_1,
\omega_2,\mathbf r, \mathbf r_{1}, \mathbf
r_{3})   \right. \nonumber \\
&-& \left. \frac{e^{2}}{2mc} \left( \sum_{\alpha } \frac{2
\Omega_\alpha j_{\alpha}^{\lambda_{s}} (\mathbf r)n_{\alpha}
(\mathbf r_{1})} {\Omega_\alpha^2 - (\omega_{1} +
\omega_{2})^2}\delta(\mathbf r_{3}
- \mathbf r_{1})\delta_{\lambda_{2} \lambda_{1}} \right. \right. \nonumber \\
&& \left. \left. + \sum_{\alpha } \frac{2 \Omega_\alpha n_{\alpha}
(\mathbf r)n(\mathbf r_{3})} {\Omega_\alpha^2 - (\omega_{1} +
\omega_{2})^2}\delta(\mathbf r - \mathbf r_{3})\delta_{
\lambda_{2} \lambda_{s}}\right)\right] ,
\end{eqnarray}
where
\begin{eqnarray}\label{paramagnetic11}
&& \hat{\chi}_{\lambda_{1} \lambda_{2}\lambda_{s}
}^{(2)}(\omega_1, \omega_2,\mathbf r, \mathbf r_{1}, \mathbf
r_{3}) = - 2 \sum_{\alpha \lambda_{s}\beta\gamma}
\frac{V_{g(-\alpha
\beta\gamma)}(\mathbf{r},\mathbf{r}_{1},\mathbf{r}_{3})j_{\alpha}^{\lambda_{s}}(\mathbf
r) n_{-\beta}(\mathbf r_{3}) j_{-\gamma}^{\lambda_{1} }(\mathbf
r_{1}) s_{\alpha } s_{\beta}}{(\Omega_{\alpha } - \omega_1
- \omega_2)(\Omega_{\beta} - \omega_1)(\Omega_{\gamma} - \omega_2)} \nonumber \\
&+& \sum_{\alpha\beta} \frac{j_{-\alpha \beta}^{\lambda_{s}
}(\mathbf r)n_{\alpha}(\mathbf r_{3})
j_{-\beta}^{\lambda_{1}}(\mathbf r_{1}) s_{\alpha}
s_{\beta}}{(\Omega_{\alpha} - \omega_1 - \omega_2)(\Omega_{\beta}
- \omega_1)} + \sum_{\alpha\beta} \frac{j_{-\alpha \beta}^{
\lambda_{s} }(\mathbf r)n_{\alpha}(\mathbf r_{3})
j_{-\beta}^{\lambda_{1}}(\mathbf r_{1}) s_{\alpha}
s_{\beta}}{(\Omega_{\alpha} - \omega_1
- \omega_2)(\Omega_{\beta} - \omega_2)} \nonumber \\
&+& \sum_{\alpha \beta} \frac{j_{\alpha \beta}^{\lambda_{s}
}(\mathbf r)n_{\alpha}(\mathbf r_{3})
j_{-\beta}^{\lambda_{1}}(\mathbf r_{1}) s_{\alpha}
s_{\beta}}{(s_{\alpha} \Omega_{\alpha} - \omega_1)(s_{\beta}
\Omega_{\beta} - \omega_2)}, \hspace{1em} \alpha, \beta, \gamma =
\pm 1, \pm2, \ldots, \label{con22}
\end{eqnarray}
where $s_{\alpha} \equiv sign(\alpha)$.


\end{document}